# Evidence for ultra long range energy transfer in organic photovoltaic donor-acceptor three dimensional films

Eran Avnon[1], Ariel Epstein[1], Olga Solomeshch[1], and Nir Tessler[1,*]

[1] Zisapel nanoelectronics center, Department of Electrical Engineering, Technion – Israel Institute of Technology, Haifa 32000, Israel

We report an ultra long range energy transfer in a layered donor–spacer–acceptor structures, which consist of the typical organic photovoltaic material system of poly(3-hexylthiophene) (P3HT) as the donor and [6,6]-phenyl C61-butyric acid methyl ester (PCBM) as the acceptor. By varying the thicknesses of spacer and acceptor layers we show that the energy transfer is to the full volume of the acceptor and not just to its nearest interface, and we find an effective energy transfer range on the order of ~100nm. Our efforts to elucidate the origin of this process, both theoretically and experimentally, are discussed as well. Although it was recently implied that an exceptionally large energy transfer may take place in this material system, this is the first time, to the best of our knowledge, that the energy transfer mechanism is characterized in a quantitative way and compared to the common models. Our analysis offers new prospects for the familiar photovoltaic bi-layer configurations, which may be very efficient if the ultra long range energy transfer is utilized through a suitable architecture.

---

[*] Corresponding Author. E-mail: nir@ee.technion.ac.il

A key consideration in organic photovoltaic (OPV) cells is the short exciton diffusion length in organic materials which implies that only excitons generated within several nanometers from a donor-acceptor heterojunction are able to dissociate into free charges and contribute to a photocurrent. [1, 2] Consequently, bi-layer OPVs were considered inefficient due to the limited heterojunction area, and the concept of bulk heterojunction (BHJ) was introduced.[3] In the BHJ structure the donor and acceptor are intimately mixed forming nanometer-scale phase separated domains throughout the entire volume, thus, decreasing the distance an exciton needs to travel before reaching a heterojunction. Yet, the increase in exciton dissociation efficiency might come at the expense of charge transport, as the former requires nanometer domains while the latter requires continuous routes to the electrodes.[4] Although the best reported OPVs to date are based on the BHJ concept, [5] satisfying both requirements is not trivial. [6] In this context, long range resonant energy transfer can be used to increase exciton harvesting in bi-layer structures [4, 7, 8] making them competitive with BHJ OPVs while retaining the good charge transport characteristics.

Resonant energy transfer (RET) between an excited donor molecule and a ground state acceptor molecule is usually considered to be an efficient process only when the donor and acceptor are in relatively close proximity, several nanometers at most. This is based on the Förster theory [9] that quantum mechanically describes the interaction between two isolated point dipoles separated by a distance $r$. It was shown that the rate of energy transfer in such case, $k_{FRET}$, can be concisely written as [10]:

$$(1) \quad k_{FRET} = \frac{1}{\tau_D^0} \cdot \left(\frac{R_0}{r}\right)^6$$

where $\tau_D^0$ is the donor excited-state lifetime in the absence of acceptor and $R_0$, usually termed Förster radius, is the distance between the donor and acceptor dipoles at which the decay rate due to energy transfer equals the donor's decay rate in the absence of the acceptor. When the Förster radius is only a few nanometers, as in many molecular systems, [11, 12] the $1/r^6$ dependency dictates that the Förster energy transfer (FRET) will be significant for short distances only. While the Förster theory was successfully used to describe RET in many useful applications [11, 13-16] there are examples where it fails to accurately predict the RET process. [17-19] One of those examples is when the point-to-point scenario does not truly represent the system used as in the case of layered thin films where an excited donor can basically interact with any of the dipoles in the acceptor layer.[4] At least intuitively, the reduction in problem dimensions implies that the $1/r^6$ dependency will reduced to $1/r^4$ in the case of a monolayer of acceptors or $1/r^3$ in the case of an infinite slab of acceptors. Such reduction in the power of $r$ is expected to be manifested in a longer energy transfer range than predicted by the Förster theory. Although the theoretical ground for the point-to-layer scenario is known for years [20, 21] this phenomenon received little attention until recently, when it was experimentally shown to be pronounced in a number of donor-acceptor systems.[4, 8, 14, 22-27]

In a recent study, Coffey *et al.*[8] have examined layered structures of donor/ electron blocking layer (spacer)/ acceptor. The electron blocking spacer was introduced in order to exclude conventional routes for exciton dissociation, i.e. exciton diffusion and dissociation at the donor/ spacer interface or short range RET. Using time-resolved microwave conductivity (TRMC) they observed, upon photoexcitation of the donor, charge generation in the spacer/ acceptor interface even in the presence of a relatively thick spacer (17nm). While these results are a clear indication that long range RET is taking place, they, as well as the cell's external quantum efficiency measurements presented in their work, are not suitable for quantitative estimation of the energy transfer rate or related parameters such as a characteristic distance. Probing the energy transfer using charge generation at the interface may ignore a large portion of the transferred energy in case it is transferred to the bulk of the acceptor film. Indeed, preliminary study by Ayzner *et. al.*[7] suggest that in the P3HT-PCBM system the transfer range may be significantly larger.

**Results**

In this contribution we systematically explore the long range RET mechanism by performing steady state photoluminescence (PL) measurements on layered donor/ insulating spacer/ acceptor structure (Illustrated in Figure 1). By varying the thickness of the spacer as well as of the acceptor layer we find that the energy transfer is 3D like, i.e. to the whole acceptor volume. We first chose regio-regular poly(3-hexylthiophene) (RR-P3HT) as the donor and [6,6]-phenyl C61-butyric acid methyl ester (PCBM) as the acceptor, materials combination frequently used in bi-layer and BHJ OPVs. In addition, we considered regio-random P3HT (RRa-P3HT) as another donor material which is blue shifted with respect to the regio-regular and has a very different morphology (PL spectra of both P3HT types are presented in Figure 2). By performing the PL measurements in an integrating sphere we were able to determine the PL efficiency (PLE) of the samples and therefore directly investigate the effect of spacer thickness, acceptor thickness, and donor PLE. We also apply a fitting procedure of our experimental data to a model describing the point dipole – finite slab scenario which allows us to extract a critical distance for the process.

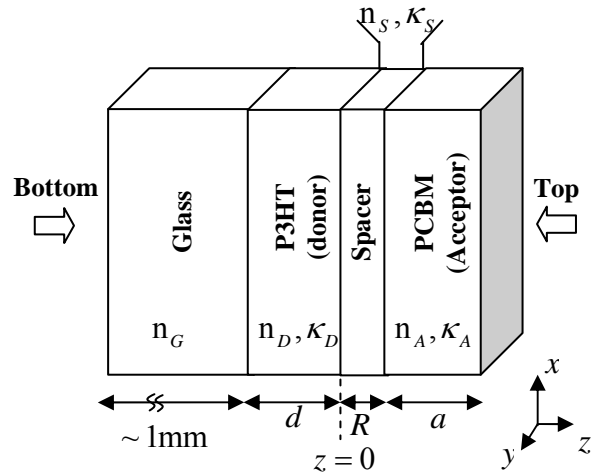

**Figure 1**. *Illustration of the layered structure of the samples examined in this work. n is the refractive index, $\kappa$ is the extinction coefficient, d is the donor thickness, R is the spacer thickness, and a the acceptor thickness. The subscripts G, D, S, and A indicate the glass, donor, spacer, and acceptor, respectively.*

Prior to describing the results we outline our processing and measurement procedures. All samples used in our experiments were fabricated in a glove box under inert atmosphere and following the same procedures. First, P3HT solution was spin coated on a clean glass substrate. Secondly, a double-layer spacer was deposited consisting of 10nm of thermally evaporated LiF and spin coated poly(vinylpyrrolidone) (PVP). The LiF thin layer serves two purposes: it is an isolating layer, and it promotes better wetting of the PVP solution allowing us to achieve highly uniform and conformal PVP layers. PVP is readily soluble in methanol therefore we could deposit it on top of the P3HT/LiF samples with no influence to the P3HT. Samples were completed with the deposition of PCBM layer (see Figure 2 for its normalized absorption spectrum). Cyclohexane was used to dissolve PCBM as it does not dissolve PVP (which is soluble in polar organic solvents) or P3HT. Yet, it is also not a very good solvent for PCBM, forcing us to work with low concentration solutions that even under low spin speed yielded no more than 30nm films.

In order to ensure none of the fabrication steps by themselves, beside of course PCBM deposition, affect in any way the absorption or PLE of the P3HT samples, we performed the measurements after each step and used control experiments where only pure solvents (methanol or cyclohexane) were spin cast. To avoid degradation of the samples due to exposure to ambient atmosphere, samples were sealed in a specially designed quartz cuvette before they were taken out of the glove-box. We found samples to be stable for days when kept inside the sealed cuvette. This procedure enabled us not only to verify that none of the fabrication steps influence the results in any way, but also to minimize experimental errors as every sample was measured and analyzed relatively to baseline measurements taken from the same sample in former steps of the fabrication process.

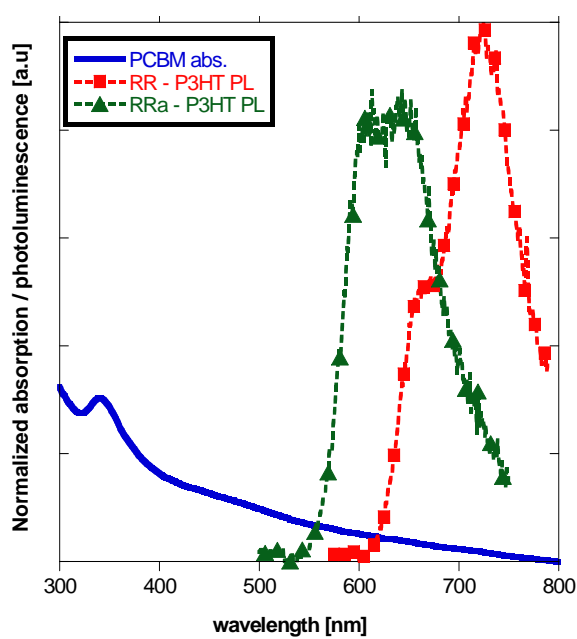

**Figure 2.** *Absorption spectrum of PCBM (full line) and photoluminescence (PL) spectra (dashed line) of regio-regular P3HT (RR-P3HT, squares) and regio-random P3HT (RRa-P3HT, triangles).*

The validity of our results strongly depends on the ability of our spacer to prevent PCBM from penetrating into the P3HT, as even small quantities of PCBM can induce pronounced quenching of the PL when blended with RR-P3HT [28] and RRa-P3HT

(Figure 3A). Therefore, the integrity of our spacer had to be thoroughly examined. Given the relatively large area of the samples (>1cm$^2$) a single characterization technique can not undoubtedly role out the existence of scarce defects, hence, it is the combined results from surface, optical and electrical characterization techniques that convinced us that our spacer contains negligible amount of defects and that PCBM do not penetrate into the P3HT. Two types of experiments we find to support these conclusions most convincingly. The first is that we found the process to be reversible. Figure 3B presents the PL spectra of the same sample at different processing stages, prior to PCBM deposition, with PCBM, and after PCBM was removed (together with the PVP) by simply spin coating the sample with a pure solvent (methanol). The pronounced quenching of PL intensity following PCBM deposition is clearly observable but also the complete recovery of the PL intensity when PCBM is washed away. It is hard to believe that by simply spin coating we could extract PCBM that penetrated into the P3HT, and the results evidently show that even if it did, the quantities were too small to influence P3HT PL.

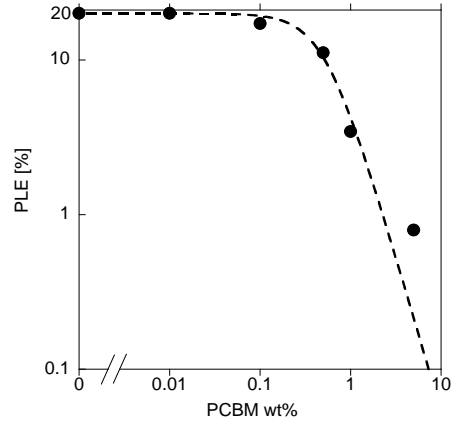

**(A)**

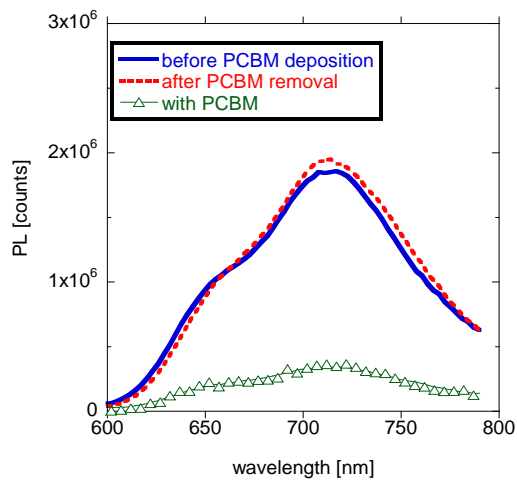

**(B)**

**Figure 3.** *(A) PLE of blend films of RRa-P3HT:PCBM for different PCBM concentrations (wt%). (B) Photoluminescence (PL) spectra of a RR-P3HT sample before PCBM deposition (full line), with PCBM (triangles), and after PCBM was removed (dashed line).*

The second experiment we performed to verify spacer integrity consisted of I-V measurements on diode structures with and without the spacer layer (but otherwise identical). We have used the same fabrication procedures and the same layer thickness as in the PLE measurements. The diodes with a spacer layer displayed in about 50% of the cases an open-circuit behavior until relatively high voltages (~8V) while in the remaining 50% we observed at least twice the turn-on voltage and more than three orders of magnitude smaller currents than those measured for the diodes without a spacer. Given the thin thickness of our spacer we find these results a good evidence for the spacer integrity suggesting that only a negligible amount of defects might exist.

After outlining the measurement and processing procedure we move to describing the data collected and its analysis. As already mentioned, PL measurements were performed in an integrating sphere and the PLE was extracted from those measurements following the procedures described by de Mello *et al.* [29]. In the absence of acceptor layer (PCBM), photogenerated excitons in the donor (P3HT) can undergo decay to the ground state through radiative and non-radiative transitions, with decay rates $k_{rad}$ and $k_{non\_rad}$, respectively. The measured PLE in this case ($\eta_0$) is given by:

$$(2) \quad \eta_0 \equiv \frac{k_{rad}}{k_{rad} + k_{non\_rad}} = \frac{k_{rad}}{k_0}$$

When the acceptor layer is introduced, another process leading to excitation quenching is enabled, namely RET from donor to acceptor, with a decay rate defined as $k_{ET}$. Therefore, the measured PLE ($\eta$) is now given by:

$$(3) \quad \eta \equiv \frac{k_{rad}}{k_{rad} + k_{non\_rad} + k_{ET}} = \frac{k_{rad}}{k_0 + k_{ET}}$$

Utilizing the reasonable assumption that $k_0$, the sum of radiative and non radiative decay rates, does not change with the introduction of the acceptor layer we can write a normalized expression for the decay rate due to energy transfer:

$$(4) \quad \frac{k_{ET}}{k_0} = \frac{\eta_0}{\eta} - 1$$

We can therefore plot $k_{ET}/k_0$ using the PLE values extracted from the measurements and investigate the influence of different parameters on the energy transfer decay rate. We begin by examining the effect of spacer thickness from a set of samples with a varying LiF+PVP thicknesses (dictated by the PVP thickness) separating a 130nm RR-P3HT layer from a 20nm PCBM layer. We observe (Figure 4A) reduction in the energy transfer rate with increasing spacer thickness, a similar behavior to that reported by others [8]. But, what is more interesting to note is that even for a relatively thick spacer (40nm) the energy transfer rate is still substantial. Using the data collected from the measurements and accounting for optical interferences we were able to simulate the exciton distribution profile, *f(z')*, within the P3HT (this procedure is described in more details further on in the paper). This profile for the case of 40nm spacer is presented in Figure 4B showing that most of the excitons are generated far from the spacer. Hence, we can conclude that the range of energy transfer may be as high as ~100nm. Such large range definitely cannot be explained using classical FRET.

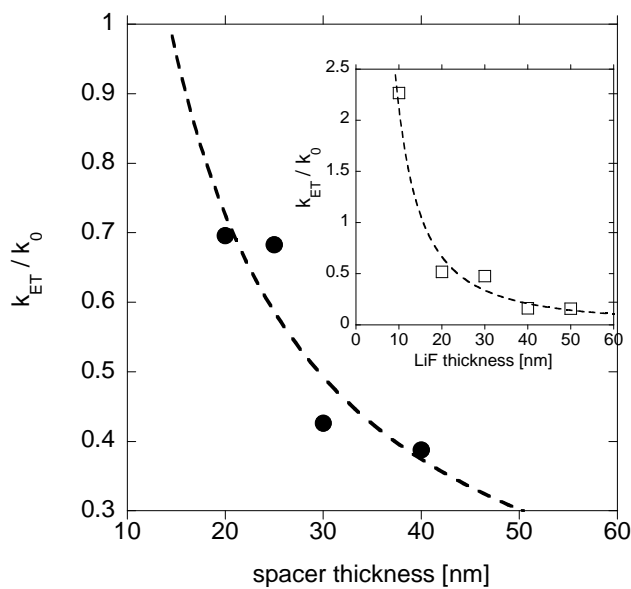

**(A)**

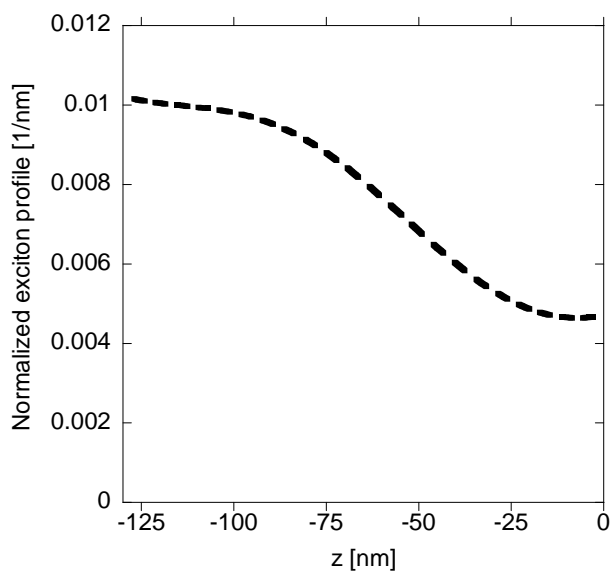

**(B)**

**Figure 4.** *(A) Normalized energy transfer rate as a function of spacer thickness. The spacer is composed of 10nm LiF and 10-30nm of PVP. P3HT and PCBM thicknesses were in all samples 130nm and ~20nm, respectively. The inset display the results for a spacer composed entirely from LiF (see text for details). Dashed lines are a guide to the eye. (B) Normalized exciton distribution profile within the P3HT layer for the case of 40nm spacer thickness.*

The inset of Figure 4A presents the results of a similar experiment where the spacer layer was constructed entirely from evaporated LiF. Comparing the two sets of results one can see that there is a rather good agreement for the range where the two overlap (20-40nm). However, as the LiF spacer on its own is less reliable than the double layer it was not used in further experiments.

The above results might suggest that in principal bi-layer OPVs should be very efficient as the long range energy transfer can bypass the limitation imposed by the exciton diffusion length. However, this is usually not the case and a possible explanation could be that excitons undergoing RET from P3HT to PCBM do not necessarily transfer to the vicinity of the heterojunction. If this is the case, what limits the efficiency is not exciton diffusion in P3HT but rather the exciton diffusion in PCBM. In Figure 5 the normalized energy transfer rate as a function of PCBM thickness is presented. Full and dashed lines are fits to measured data using the model described later in the text (Eq. (10)). In this experiment the P3HT and LiF+PVP spacer thicknesses were fixed to 130nm and ~20nm, respectively, and we examined both the cases of bottom facet illumination and top facet illumination (see Figure 1).

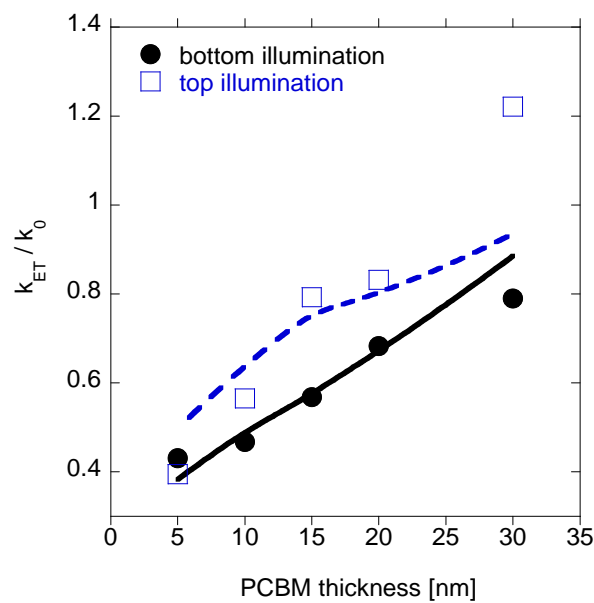

**Figure 5.** *Normalized energy transfer rate as a function of PCBM thickness for samples based on RR-P3HT. RR-P3HT thickness is 130nm and spacer (LiF+PVP) is ~20nm thick. Filled circles correspond to illumination from the bottom facet (P3HT side) and squares correspond to illumination from the top facet (PCBM side). Full and dashed lines are fits to measured data using the model described in the text below for the bottom and top illumination, respectively.*

The results clearly indicate that energy transfer rate is increasing with PCBM thickness. This trend cannot be explained by optical interference phenomena according to our optical simulation (not shown). Another interesting point arising from Figure 5 is that the differences between the bottom and top illumination directions are relatively small (with the exception of the thickest PCBM considered). This is somehow in contradiction to one's expectations, as in the case of top illumination more excitons should be generated close to the P3HT/spacer interface, which is expected to significantly enhance energy transfer. However, considering optical interferences and especially the large range RET we observe (effectively ~100nm), which is not very far from the P3HT film thickness (130nm), such small differences are reasonable and even serve as another indication for the ultra long range RET.

After we have examined the influence of structural parameters (spacer and acceptor thickness) we turned to investigate material parameters that might shed light on the origin of this long range RET. We chose regio-random P3HT (RRa-P3HT) for this purpose for two reasons. Firstly, we have measured its PLE to be an order of magnitude higher than that of the RR-P3HT (20% compared to 2%), while the spectral overlap integrals are essentially the same (calculated following Kuhn [30]). An increase in the donor PLE is expected to be manifested in a higher rate of energy transfer according to the Förster theory.[30] Secondly, the degree of regio-regularity in P3HT was found to influence the absorption and PL spectra of this material due to coherence effects related to interchain excitons [31, 32]. Such effects were predicted to lead to deviations from the Förster theory in some systems [18].

To this end we have fabricated samples based on RRa-P3HT in similar fashion to those made from RR-P3HT. In Figure 6 the normalized energy transfer rate as a function of PCBM thickness is presented for these samples. The energy transfer rate practically follows the same trend as in the case of RR-P3HT, i.e. increasing with PCBM thickness. The difference between the two materials is in the magnitude of the normalized energy transfer rate, where RRa-P3HT displays an order of magnitude higher normalized transfer rates. The fact that the increase in the energy transfer rate corresponds almost perfectly to the increase in PLE suggests that this factor is probably due to lower non-radiative rate in RRa-P3HT (which leads to lower normalization factor, $k_0$).

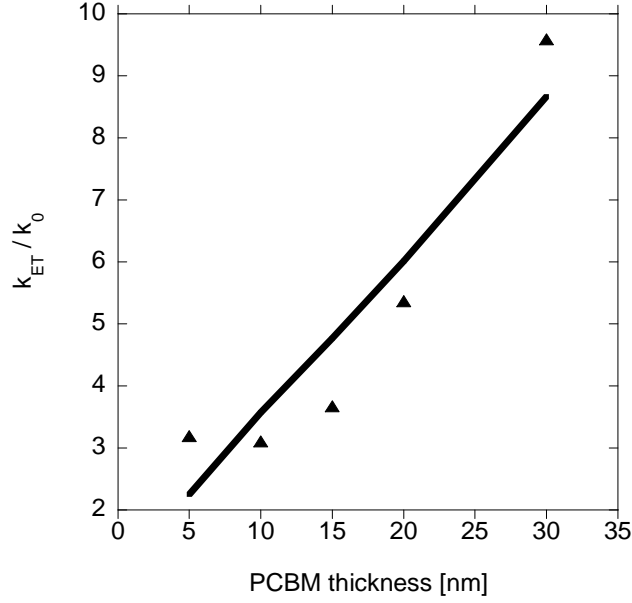

**Figure 6.** *Normalized energy transfer rate as a function of PCBM thickness for samples made from RRa-P3HT (130nm). Spacer (LiF+PVP) thickness is ~25nm. Full line is a fit to measured data using the model described in text.*

**Modeling and discussion of energy transfer in stratified media**

We emphasize that all the above observations do not rely on any model but are solely based on the definitions of PLE (Eqs. 2, 3, and 4) and on the values of this parameter as were extracted from our measurements. Still, fitting the experimental data to a model can help elucidating the origin of this RET process.

Considering a theoretical modeling for the energy transfer phenomenon, it was already shown that the results obtained by Förster quantum mechanically can be reproduced using classical electromagnetic theory [20, 21, 30]. In the case of point-to-point dipole model, the decay rate due to energy transfer between a donor dipole located at $\vec{r}_D$ and an acceptor dipole located at $\vec{r}_A$, is given, very similarly to Eq.(1), by

$$(5) \quad \frac{k_{ET}(r)}{k_0} = \left(\frac{R_0}{r}\right)^6$$

where $\vec{r} = \vec{r}_D - \vec{r}_A$ and the Förster radius is defined as

$$(6) \quad R_0 = \left(\frac{3}{128\pi^5} \frac{K^2 \alpha(\lambda_0) \eta_0 \lambda_0^4}{C_A n^4}\right)^{1/6}$$

where $\lambda_0$ is the wavelength of the oscillating donor dipole in free-space, n is the donor refractive index (assuming the donor and acceptor refractive indices are similar), $\alpha(\lambda_0)$ is the wavelength-dependent absorption coefficient (cm$^{-1}$) of a film made of the acceptor molecules, and $C_A$ [cm$^{-3}$] is the density of acceptor dipoles in that film (i.e. $\alpha(\lambda_0)/C_A$ is the absorption cross section of a single molecule). Assuming the donor dipoles are aligned in-plane and the acceptor is isotropic (a reasonable assumption when considering the donor and acceptor molecular structure), the orientation factor, K, is given by $K^2 = 1 + 3\cos^2\theta$, where $\theta$ is the angle between the donor dipole moment axis and the direction of $\vec{r}$.

As we already briefly pointed out, the stratified structures used in our experiments do not necessarily match the physical picture underlying the Förster theory (point dipole-to-point dipole). Following previous works, [8, 21, 24-26] we extend the formulation presented in Eqs. (5) and (6) to a film-to-film scenario, taking into account both the acceptor finite thickness and the exciton distribution within the donor layer. We consider a layer of generally uncorrelated donor molecules, separated by a spacer from a layer of uncorrelated acceptor molecules (i.e. we ignore potential coherence effects), all fabricated on a thick glass substrate as illustrated in Figure 1. We denote the donor, spacer and acceptor thicknesses along the z-axis as $d$, $R$ and $a$, respectively, and the other two dimensions are considered to be infinite. The layers are characterized optically by their refractive index, n, and extinction coefficient, $\kappa$. The wavelength dependent values of the refractive index were taken from the literature [33], while the extinction coefficients were extracted from our absorption measurements. We assume that the spacer is not interacting with the dipoles neither of the donor molecules, nor of the acceptor molecules, and that there is some overlap between the donor luminescence spectrum and the acceptor absorption spectrum to allow an efficient dipole-dipole interaction between the two species.

Given an excited dipole located at a certain point, $z = z'$, in the donor layer and a homogenous acceptor dipole density, $C_A$, the energy transfer rate expression in Eq. (5) can be generalized to fit the dipole to layer scenario [21]:

$$(7) \quad \frac{k_{ET}(z')}{k_0} = d_0^{\ 3}\left[\frac{1}{(R-z')^3} - \frac{1}{(R+a-z')^3}\right]$$

where $d_0$ is known as the critical length of the energy transfer process. [21, 26] By rigorously deriving the above equation we find that $d_0$ can be correlated with the Förster radius, and through it to material parameters, according to

$$(8) \quad d_0 = \left( \frac{\pi C_A}{4K^2} \right)^{1/3} R_0^2$$

We note that once more we assumed the donor dipoles to be aligned in-plane and the acceptor to be isotropic, and that $z' < 0$ with accordance to Figure 1.

We take into account the distribution of donor dipoles in our model by introducing the photo-generated exciton distribution function, $f(z')$, which is proportional to the photon absorption profile resulting from the monochromatic illumination. While in the absence of acceptor layer the PLE is considered independent of the exciton position within the layer, when the acceptor is introduced and the energy transfer process is enabled, $f(z')$ should be accounted for. Therefore, using $f(z')$ as a normalized probability density function, we rewrite Eq. (3) to represent the ratio between the number of emitted photons from an interval $dz'$ in the donor layer to the total number of absorbed photons:

$$(9) \quad d\eta(z') = \frac{k_{rad}}{k_0 + k_{ET}(z')} f(z') dz'$$

Integrating over the donor layer we arrive at an expression for the measured PLE in the presence of the acceptor:

$$(10) \quad \eta = \eta_0 \int_{-d}^{0} \frac{f(z')}{1 + d_0^3 \left[ (R - z')^{-3} - (R + a - z')^{-3} \right]} dz'$$

We note that all the parameters in Eq. (10) can be measured experimentally except $f(z')$ and $d_0$. As $f(z')$ follows the spatial absorption profile, it can be readily calculated by a transfer-matrix or recursive approach for the stratified device we examine, using the optical properties of the media at the excitation wavelength. [34] As the measurements were done inside an integrating sphere one has also to account for indirect excitation [29] which in our case modified slightly the calculated exciton profile. Having calculated the exciton profile (as in Figure 4B) we are left with a single parameter, $d_0$, which can be used to fit the model to the experimental data.

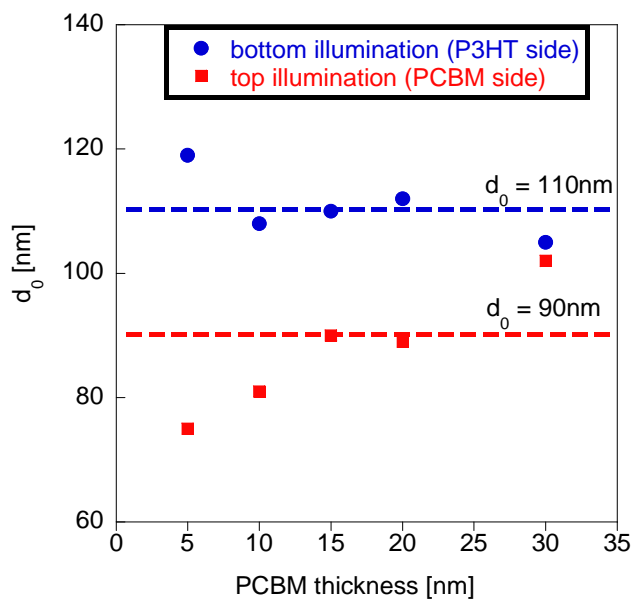

**(A)**

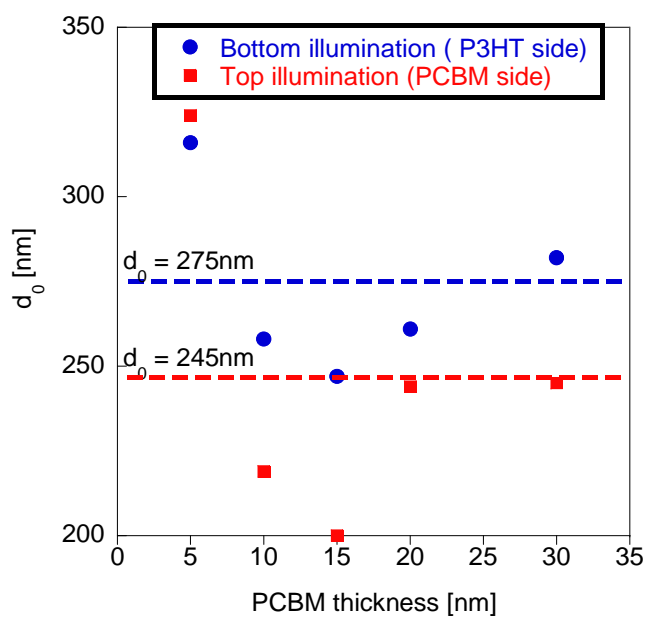

**(B)**

**Figure 7.** *Values of $d_0$ extracted using the model described in the text for the experiments where PCBM thickness was varied. (A) RR-P3HT for both top and bottom illumination. (B) RRa-P3HT for bottom illumination. Dashed lines are to guide the eye.*

Values extracted for $d_0$ using the above model are presented in Figure 7 for samples based on RR-P3HT (Figure 7A) and RRa-P3HT (Figure 7B). We obtained an average values of ~110nm (±5%), ~90nm (±12%), ~275nm (±10%), 245nm (±20%) for bottom illuminated RR-P3HT, top illuminated RR-P3HT, bottom illuminated RRa-P3HT, and top illuminated RRa-P3HT, respectively. With the exception of the top-illuminated RRa-P3HT, the relatively small deviations are within reasonable experimental errors. The slight difference between $d_0$ extracted for top and bottom illumination is most likely due to the exciton profile being more uniform across the sample than our optical modeling suggests. The lines drawn in Figure 5 and Figure 6 are based on Eq. (7), using the average values of the above extracted $d_0$, providing further support of the analysis procedure.

However, a more detailed examination of the values extracted for $d_0$ reveals that this model may be too simplified to accurately describe this ultra long range RET. Although $d_0$ does not hold the same meaning as the Förster radius ($R_0$), one can be calculated from the other according to Eq. (8). Using the values we have extracted for $d_0$ and considering $C_A$ ~2 x $10^{21}$cm$^{-3}$ and $K^2$ ~1 we find the corresponding $R_0$ to be ~9 nm for RR-P3HT and ~14nm for RRa-P3HT. Yet, $R_0$ values calculated using Eq. (6) in conjunction with the data shown in Figure 2 were found to be ~1.2nm for RR-P3HT and ~1.8nm for RRa-P3HT.

Considering the above, we may deduce that the ultra long range energy transfer is not merely an extension of the Förster theory to the case of 3D layers, and other physical phenomena, not taken into account in the current model, should be introduced to the model. We have extended the oscillating dipole model to include the full dipole electrical field terms (e.g., far field interaction), yet since the acceptor (PCBM) layer is very thin this had rather small effect. Another direction that we considered was solving numerically the exciton transport equation in the P3HT layer in the presence of generation/decay, diffusion, and energy transfer processes in order to get a more accurate evaluation of the exciton profile. The results showed that the exciton diffusion had small effect on the resulting $d_0$ (assuming exciton diffusion length in P3HT is not more than 10nm). The two effects together could potentially reduce $d_0$ by up to ~30% which is not sufficient to account for the difference in $R_0$ values. Another process which is overlooked by the current model is related to correlations between adjacent donor dipoles; large $d_0$ values may imply that the donor excitations are not singular, but rather form some kind of a collective coherent source spread over several excited states, which may result in a different field dependency, thus increasing the effectiveness of energy transfer. The possibility for such a phenomenon to occur in organic semiconductors has recently been pointed out by Emelianova *et. al.* . [35] We believe that the fact that the difference between regio-regular and regio-random P3HT could be accounted for solely by the different PLE values lowers the probability for some of the potential coherent-configurations to hold but does not rule out the concept. Additional possible mechanisms may arise from other quantum effects [18, 19] requiring different approach than the common classical oscillating dipole model.

**Conclusions**

In conclusion, a clear evidence for the existence of an ultra long range resonant energy transfer in a stratified structure has been presented, reaching an effective transfer range of over 100nm. By photoluminescence efficiency measurements, we have shown this process to depend on the acceptor and spacer layers thicknesses indicating it to be a 3D process. While the fact that the extracted $d_0$ values were found to follow the donor PLE supports the approach based on a Coulomb-like interactions, the common classical oscillating dipole model, extended to suit the experiment conditions, supplies only partial explanations for this phenomenon. In future investigations we intend to extend the model to account for extended donor sources, which we believe could shed light on the physical origins of the presented ultra long range energy transfer. The implications of our finding that the transfer is to the full volume of the acceptor are twofold. Firstly, this may provide an alternative explanation for the report in ref [7] that, for a bi-layer device, when a thicker donor material is used it is also important to use a thicker acceptor layer. In the current context this is so to improve the energy transfer. Secondly, a design of bi-layer structure should take into account that some of the excitons that are transferred to the acceptor will be generated at the top interface and hence it would be essential to add an exciton blocking layer that would prevent their quenching by the metal electrode.

## Methods

Regio-regular and regio-random P3HT were purchased from Rieke metals, PCBM was purchased from Nano-C. PVP was purchased from Sigma-Aldrich. All solvents used were anhydrous (Sigma-Aldrich). All materials and solvents were used as received with no further purification.

Glass substrates (> 1cm x 1cm) were cleaned in an ultrasonic bath with water, acetone, methanol and 2-propanol consecutively. Substrates were then dried on a hot plate and UV-ozone treated for 15 minutes before they were immediately transferred to the glove box.

P3HT (both types) solutions in chloroform were prepared and spin coated on the glass substrate followed by annealing in a vacuum oven ($110^0$C, 3hrs.). LiF was thermally evaporated in a rate of 0.5-1Å/sec with base pressure of at least $1\cdot10^{-6}$ mbar. A 1wt% solution of PVP in methanol was prepared and spin coated on the sample following the LiF evaporation and then dried in a vacuum oven ($70^0$C, 2hrs.). PCBM was dissolved in hot cyclohexane at a maximum concentration of ~1mg/ml, solutions were continuously stirred on a hot-plate for several days. After spin coating the samples with PCBM they were dried in a vacuum oven at room temperature for at least 12hrs.

All solutions were filtered prior to spin coating using a 0.2μm PTFE filter. Films thicknesses were evaluated using a surface profiler (Dektak 150, Veeco). And the films surface was examined using AFM (Dimension 3000, Digital Instruments). Absorption spectra were recorded using UV-VIS-NIR scanning spectrophotometer (UV-3101PC, SHIMADZU).

PLE measurements were performed in an integrating sphere (labsphere) that was coupled using fiber optic cables to an Edinburgh spectrometer system equipped with a single photon photomultiplier (S300) detector. For the RR-P3HT excitation a 532nm CW laser was coupled to inlet fiber optic cable, and for RRa-P3HT excitation we used the system Xe arc lamp and monochromator to excite in a 470nm wavelength.

The optical constants of the various layers with which the exciton distribution was calculated are, in correspondence to Fig. 1: $n_G$=1.53, $n_S$=1.39, $\kappa_S$=0. For RR-P3HT at 532nm wavelength: $n_D$=1.945, $\kappa_D$=0.3. For RRa-P3HT at 470nm wavelength: $n_D$=1.8, $\kappa_D$=0.15. For PCBM: $n_A$(@470nm)=2.24, $\kappa_A$(@470nm)=0.2, $n_A$(@532nm)=2.1795, $\kappa_A$(@532nm)=0.114.

# Acknowledgements

We acknowledge the support of the Russell Berrie Nanotechnology Institute at the Technion. E.A acknowledges the support of the Israeli ministry of science. We also acknowledge fruitful discussions with Evguenia Emelianova.